\documentclass[journal]{IEEEtran}

\usepackage[noadjust]{cite}

\ifCLASSINFOpdf
\else
\fi

\usepackage[cmex10]{amsmath}
\usepackage{array}
\usepackage{amssymb}
\usepackage{multirow}

\usepackage[dvipsnames]{xcolor}

\newskip\smallskipamount 
\def\smallskip{\vskip\smallskipamount}

\usepackage[usenames,dvipsnames]{pstricks}
\usepackage{epsfig}
\usepackage{pst-grad} 
\usepackage{pst-plot} 
\usepackage{pst-node} 
\usepackage{threeparttable}

\usepackage{pst-circ} 

\usepackage{graphicx}
\usepackage{subcaption}
\usepackage{caption}

\usepackage[ruled,norelsize]{algorithm2e}

\makeatletter
\newcommand{\removelatexerror}{\let\@latex@error\@gobble}
\makeatother

\begin{document}
\title{A Complexity Reduction Method for Successive Cancellation List Decoding}
\author{Onur~Dizdar}
\maketitle

\begin{abstract}
This brief introduces a hardware complexity reduction method for successive cancellation list (SCL) decoders. Specifically, we propose to use a sorting scheme so that $L$ paths with smallest path metrics are also sorted according to their path indexes for path pruning. 
We prove that such sorting scheme reduces the input number of multiplexers in any hardware implementation of SCL decoding from $L$ to $(L/2+1)$ without any changes in the decoding latency. We also propose sorter architectures for the proposed sorting method. Field programmable gate array (FPGA) implementations show that the proposed method achieves significant gain in hardware consumptions of SCL decoder implementations, especially for large list sizes and block lengths.
\end{abstract}

\vspace{-0.25cm}

\IEEEpeerreviewmaketitle

\section{Introduction}
\vspace{-0.15cm}
\IEEEPARstart{S}{uccessive}-Cancellation (SC) is the first decoding algorithm proposed for polar codes by Ar\i kan in \cite{arikan}. Being a low-complexity algorithm, SC brings a penalty in the achievable error performance. In \cite{tal_list}, successive cancellation list (SCL) decoding was proposed to improve the error performance, following similar ideas in \cite{DumerList} developed for Reed-Muller (RM) codes. 

\nocite{stimming_llr, stimming_arch, yuan_multibit, fan_doublethr, fan_lowlatency, hashemi_parititoned, hashemi_sphere, hashemi_fast, sural_thesis, song_efficient, zhang_hardware, lin_hightp, xiong_multimode, xia_onpath, xiong_emulation, lin_efficient, xiong_symbol}
A common problem of SCL decoder implementations is the high hardware complexity, which is mostly due to memory elements and large multiplexers in the designs. 
Memory elements in SCL decoder implementations store calculated log-likelihood ratio (LLR) values, decoded bits, partial-sums and pointers for each path. 

Multiplexers are used to copy the memory contents between decoding paths after path pruning stages. $L$-to-$1$ multiplexers are used for this purpose for each of $L$ paths, a structure which is commonly referred as a \textsl{crossbar}. Input widths of these multiplexers are equal to the widths of the corresponding memory elements. 
For example, in SCL decoder architectures \cite{stimming_llr}-\cite{zhang_hardware}, $L$-to-$1$ multiplexers with input widths of $\mathcal{O}($N$)$ bits (\textsl{e.g.} $N$ bits, $N/2$ bits, etc.) are used to copy the contents of registers storing decoded bits and/or partial-sums between paths. 
In the cases where random access memory (RAM) blocks are used for storage, which is the conventional approach for storing calculated LLR values, pointer memories are used \cite{stimming_arch}. 
In such cases, $L$-to-$1$ multiplexers with input widths equal to the width of pointer registers are required for each path, the widths being $\mathcal{O}(\log_{2} N\log_{2} L)$ bits.  
Similar arguments are valid for designs \cite{lin_hightp}-\cite{xiong_symbol}, where decoded bits and/or partial-sums are stored in RAM blocks or partly in registers and partly in RAM blocks.

In this work, we propose a method to reduce the hardware complexity of SCL decoder implementations. We achieve this reduction by limiting the number of possible interactions between decoding paths by applying a novel sorting mechanism to determine the surviving paths at decision making stages of SCL decoding. 
Specifically, $L$ \textsl{surviving} paths, which are chosen out of $2L$ \textsl{candidate} paths, are obtained as sorted with respect to their path indexes and path copying operations are performed according to the result of this sorting mechanism. 
We prove that the proposed method reduces the input number of memory copying multiplexers from $L$ to $(L/2+1)$.
We also describe sorter architectures to enable the proposed method in an SCL decoder. 

The rest of the paper is organized as follows. We give background information on SCL decoding in Section~\ref{sec:background}. The proposed complexity reduction method and sorter architectures are described in Section~\ref{sec:method}. Section~\ref{sec:results} gives the implementation results. Section~\ref{sec:conclusion} concludes the paper.

\begin{figure*}[h!]
\vspace{-0.3cm}
\hspace{1.4cm}\includegraphics[width=6.0in,height=6.0in,keepaspectratio]{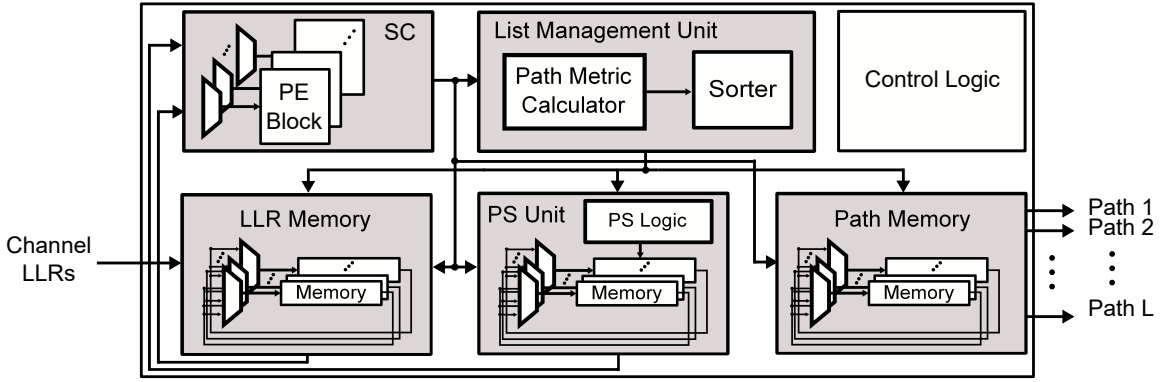}
\caption{Generic architecture for SCL decoders} 
\vspace{-0.5cm}
\label{fig:scl_arch}
\end{figure*} 
\section{Successive-Cancellation List Decoding}
\label{sec:background}
\vspace{-0.1cm}
Vectors are denoted by bold lowercase letters. We use $\mathbf{c}_{i}^{k}$ to denote the vector \mbox{$(c_{i}, c_{i+1}, \ldots, c_{k})$}.
For any set $\mathcal{S} \subseteq \left\{0,1,\ldots, N-1\right\}$, $\mathcal{S}^{\mathrm{c}}$ denotes its complement. 

A high level description of the SCL decoding algorithm is given in Algorithm~\ref{alg:HiLevelSCL}. The indexes of information bits in the uncoded bit vector $\mathbf{u}$ of length-$N$ are chosen from $\mathcal{A}$, which is the set of indexes of $K$ {\sl polarized channels} with smallest Bhattacharyya parameters \cite{arikan}. The probability $W_{N}^{(i)}\left(\mathbf{y}, \mathbf{u}_{0}^{i-1}[k]|u\right)$ is defined as the decision probability for the $i$-th bit of the $k$-th decoding path for the bit value $u \in \left\{0,1\right\}$.
SCL decoders keep $L$ decoded bit sequences during the decoding process in order to enhance the error performance.
Decoding paths are formed during the decision making stages of SC decoding for information bits. 
An existing path is split into two \textsl{candidate} paths for the bit decision values of $0$ and $1$ for the decoded information bit.

In order to limit the exponential growth of decoding paths, \textsl{path pruning} is performed to limit the number of paths to maximum list size $L$. When the number of candidate paths exceed $L$, an SCL decoder chooses $L$ paths as the surviving paths according to their \textsl{path metrics}. Path metrics are calculated from decision probabilities or LLR values depending on the decoder implementation. 

After path pruning, new paths are formed from existing paths. In software implementations of SCL decoding, the existing decoding paths are continued, copied or killed if one, both or none of their candidate paths survive, respectively. 
In hardware implementations, memory copying operations are required to perform such task, which require $L$-to-$1$ multiplexers for each memory element. In the next section, we describe a sorting method to reduce the number of inputs of such multiplexers down to $L/2+1$.

\section{Complexity Reduction Based on Multiplexers in SCL Decoders}
\label{sec:method}
\vspace{-0.15cm}
As mentioned in the previous chapter, an SCL decoder forms candidate paths from existing paths for each information bit.  
In this work, we assume that such operation is performed by forming candidate paths with indexes $2l-1$ and $2l$ from the existing path with index $l$, $1 \leq l \leq L$. 
Path metrics of the candidate paths can be calculated as in \cite{stimming_llr} or \cite{yuan_llr}.

A generic SCL decoder architecture is given in Fig.~\ref{fig:scl_arch}. In conventional hardware implementations of the SCL algorithm, decoding paths are assigned to dedicated hardware elements, \textsl{i.e.}, circuitry for calculations and memory elements. The dedicated circuitry of a decoding path consists of processing elements (represented by PE block in the SC module) and partial-sum logic (represented by PS logic in the PS unit) and performs SC decoding calculations for the specific path. The dedicated memory elements of a decoding path store the calculated LLR values, partial-sum bits, previously decoded bits and/or pointers to specify the RAM blocks which the particular path should access for decoding calculations.

After path pruning, the dedicated hardware of decoding paths are assigned to new surviving paths to continue with the decoding operations. The memory contents for each new surviving path are copied from the corresponding dedicated hardware of an existing path according to an ordering. The ordering is determined by a sorter or a module which finds the $L$ candidate paths with smallest path metrics. 
Then, such $L$ candidate paths are chosen as surviving paths and they are assigned to dedicated hardware in the specific order at the output of the sorter. 
The memory contents of the existing paths, which the surviving paths are formed from, are copied from their respective dedicated memory elements to the new memory elements according to this ordering. 
Similar arguments are valid for different types of modules that extract the $L$ surviving paths with smallest metrics without a specific ordering. 
\begin{figure}[hbt]
\vspace{-0.4cm}
 \removelatexerror
  \begin{algorithm}[H]
	 \caption{$\mathbf{\hat{u}}=\textsc{SCL}(\mathbf{y}, \mathcal{A}, \mathbf{u}_{\mathcal{A}^{\mathrm{c}}}, L)$}
		\label{alg:HiLevelSCL}
		\small
			$N=$\textit{length}$(\mathbf{y})$, \ $\gamma=1$ \\
			\For{$i = 0$ $\mathrm{\mathbf{to}}$ $N-1$}
			{
				\eIf{$i \notin \mathcal{A}$}{
								$\hat{u}_{i}[k] \gets u_{i}$, $\forall k \in \left\{1, \ldots, L\right\}$
				}{
					\eIf{$\gamma < L$}{
						\For{$k = 1$ $\mathrm{\mathbf{to}}$ $\gamma$}
						{
							$\mathbf{\hat{u}}_{0}^{i-1}[\left\{2k-1, 2k\right\}] \gets \hat{u}_{0}^{i-1}[k]$ \\
							$\hat{u}_{i}[2k-1] \gets 0$ \\
							$\hat{u}_{i}[2k] \gets 1$
						}
						$\gamma \gets 2\gamma$
					}{
						$\mathbf{\Gamma} \gets$\textsc{Sort}$\left(\left(\mathbf{\hat{u}}_{0}^{i-1}[k], u\right), W_{N}^{(i)}\left(\mathbf{y}_{0}^{N-1}, \mathbf{u}_{0}^{i-1}[k]|u\right)\right)$, $\forall k \in \left\{1, \ldots, L\right\}$, $\forall \hat{u} \in \left\{0,1\right\}$  \\ 
						   $\mathbf{\hat{u}}_{0}^{i}[k] \gets \mathbf{\Gamma}_{k}$, $\forall k \in \left\{1, \ldots, L\right\}$ \hspace{0.7cm}  \\
					}
				}
			}
			$k^{\prime} \gets \mathrm{argmax}_{k \in \left\{1,\ldots,\gamma\right\}}W_{N}^{(N-1)}\left(\mathbf{y}_{0}^{N-1}, \mathbf{u}_{0}^{N-2}[k]|u_{N-1}[k]\right)$ \\
			\Return $\mathbf{\hat{u}}[k^{\prime}]$
  \end{algorithm}
	\vspace{-0.5cm}
\end{figure}

The copying operations explained above are performed by crossbars consisting of $L$-to-$1$ multiplexers for each of $L$ paths, as shown in Fig.~\ref{fig:scl_arch}.
The input widths of such multiplexers are required to be $\mathcal{O}($N$)$ bits if they are employed to copy partial-sum and decoded bit registers. 
Calculated LLR values are conventionally stored in RAM blocks, as the number of bits to be stored is higher than those of partial-sums and decoded-bits.
In this case, pointer memories with input widths of $\mathcal{O}(\log_{2} N\log_{2} L)$ bits are employed to map calculated LLR RAM blocks to paths \cite{stimming_arch} and are copied by crossbars with equal widths. Similarly, in architectures where partial-sums are stored in RAM blocks or partly in registers and partly in RAM blocks, crossbars are employed to copy the pointer memories and the registers that store portions of partial-sums.
\vspace{-0.85cm}
\subsection{Proposed Method}
\label{sec:sorter}
We achieve a reduction in the input number of crossbar multiplexers in an SCL decoder. The reduction is achieved by ordering the indexes of surviving paths, so that dedicated memory elements of each decoding path can get memory contents from a limited set of other dedicated paths. We prove that the number of elements in such a set is $\left(\frac{L}{2}+1\right)$, so that the employed multiplexers are $\left(\frac{L}{2}+1\right)$-to-$1$ instead of $L$-to-$1$.

\textit{Proposition:} In an SCL decoder implementation, the number of paths that can have its memory contents copied to a specific path after path pruning is $\left(\frac{L}{2}+1\right)$ instead of $L$, if the $L$ surviving paths are ordered according to their indexes.

\textit{Proof:} We denote the candidate paths with index $i$, \mbox{$1 \leq i \leq 2L$}. A candidate path $i$ is formed from the existing path \mbox{$\left\lfloor \frac{i-1}{2}\right\rfloor+1$} according to our path splitting definition. 

After path pruning, $L$ candidate paths with indexes \mbox{$i_{1}$, $i_{2}$, $\ldots$, $i_{L}$} survive, where \mbox{$1 \leq i_{k} \leq 2L$} and \mbox{$\forall k \in \left\{1, 2, \ldots, L\right\}$}. The surviving path $i_{k}$ is assigned to path $k$ to continue with the decoding operations and the memory contents of the existing path \mbox{$\left\lfloor \frac{i_{k}-1}{2}\right\rfloor+1$} are copied to the memories of path $k$. When the surviving path indexes are sorted, they satisfy the expression
\vspace{-0.15cm}
\begin{align}
	i_{1} < i_{2} < \ldots < i_{L}.
\label{eqn:path_index_pre}
\end{align}
Since \mbox{$1 \leq i_{k} \leq 2L$}, expression~\eqref{eqn:path_index_pre} implies that surviving path indexes are limited by specific minimum and maximum values. For any $i_{k}$, the minimum and maximum values are found as  
\vspace{-0.15cm}
\begin{align}
	k \leq i_{k} \leq 2L-(L-k).
	\label{eqn:path_index}
\end{align}
In order to find all possible indexes of existing paths that the \mbox{$k$-th} surviving path can originate from, we use \eqref{eqn:path_index} to write 
\vspace{-0.15cm}
\begin{align}
	\left\lfloor \frac{k-1}{2}\right\rfloor+1 \leq \left\lfloor \frac{i_{k}-1}{2}\right\rfloor+1 \leq \left\lfloor \frac{L+k-1}{2}\right\rfloor+1.
	\label{eqn:path_index_2}
\end{align}
Expression \eqref{eqn:path_index_2} shows that dedicated memories of the \mbox{$k$-th} surviving path can get memory contents from dedicated memories of paths with indexes specified by the limits. It is straightforward to show that there are $\left(\frac{L}{2}+1\right)$ elements in the interval \eqref{eqn:path_index_2} for $L$ being an even number, which completes the proof.

\begin{figure}[h!]
	\vspace{-0.4cm}
	\hspace{-0.1cm}\includegraphics[width=3.5in,height=3.5in,keepaspectratio]{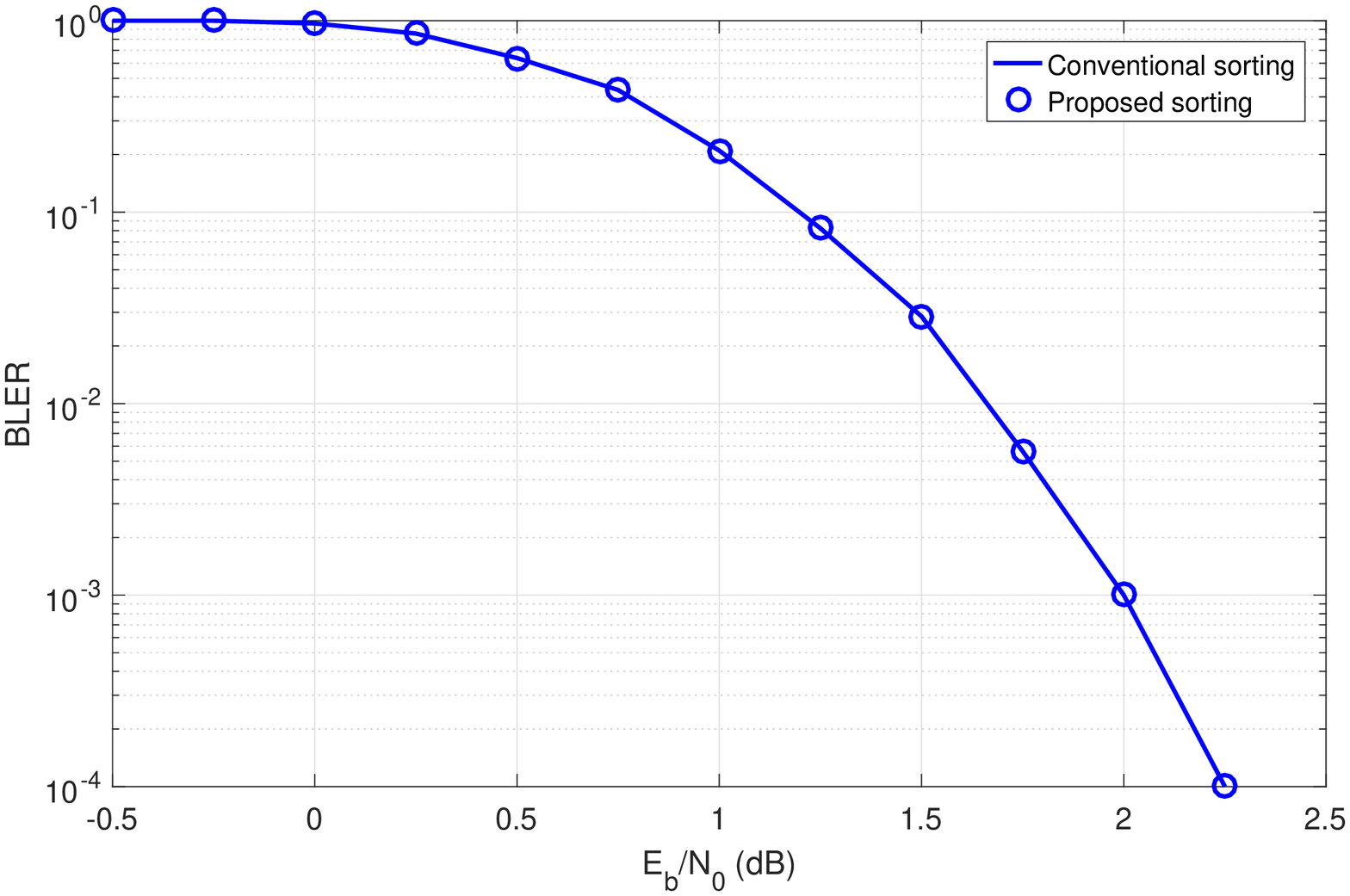}
	\caption{SCL decoding performance with conventional and propsed sorting, $N=1024$, $K=512$, $L=8$.} 
	\vspace{-0.4cm}
	\label{fig:performance}
\end{figure}

The proposed sorting method does not affect the latency or the error performance of SCL decoding. We demonstrate the error performances of SCL decoding with conventional and proposed sorting methods in Fig~\ref{fig:performance}. 
\vspace{-0.4cm}
\subsection{Sorter Design for the Proposed Method}
\label{sec:sorter}
\begin{figure}[h!]
	\hspace{0.7cm}\includegraphics[width=3.0in,height=3.0in,keepaspectratio]{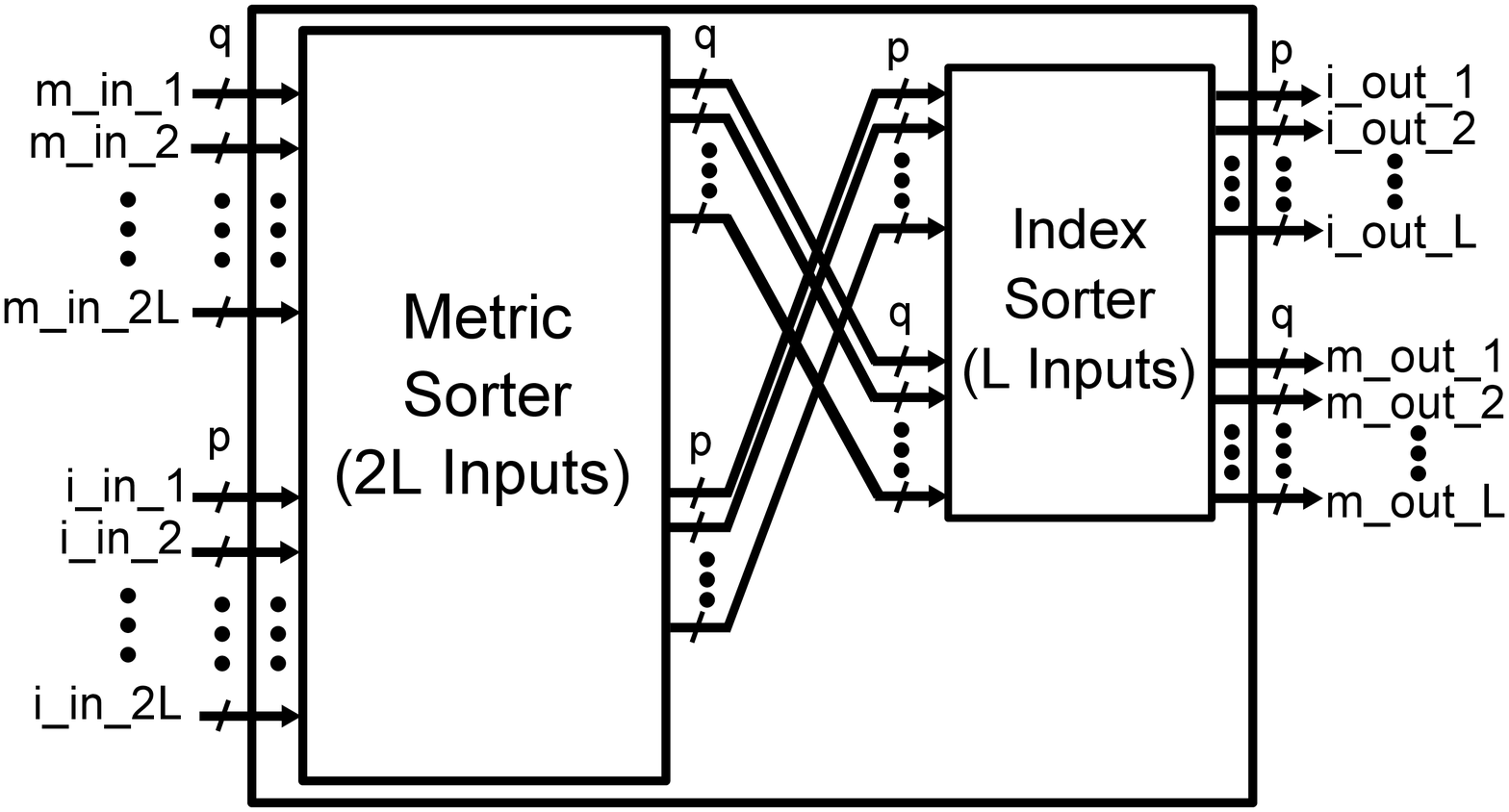}
	\caption{Proposed sorter architecture} 
	\label{fig:sorter_arch}
\vspace{-0.6cm}
\end{figure}

In this section, we give sorter designs for the proposed method. A generic sorter architecture for the method is given in Fig.~\ref{fig:sorter_arch}. The architecture takes the path metrics and path indexes of $2L$ candidate paths as inputs,denoted by \textsl{m\_in\_k} and \textsl{i\_in\_k}, respectively. We use a two stage sorter. The first stage finds the surviving paths according to their path metrics. The second stage sorts the surviving paths according to their indexes. The architecture outputs the path metrics and path indexes of $L$ surviving paths, denoted by \textsl{m\_out\_k} and \textsl{i\_out\_k}, and quantized by $q$ and $p$ bits, respectively.    

We propose $3$ different sorter designs. Our aim is to offer a trade-off between hardware complexity and throughput by the proposed designs. 
One of the sorter architectures we employ in our designs is the radix-$2L$ sorter \cite{stimming_arch}, \cite{radix_masera}.
Other sorting methods we consider are the bitonic sorter proposed in \cite{batcher_sorting} and maximum values filter (MVF) proposed in \cite{lin_efficient}. MVF is a bitonic sorter without the final sorting stages, so that it extracts $L$ paths with the smallest path metrics out of $2L$ candidate paths without any ordering. Table~\ref{table:sorter_designs} summarizes the proposed designs and their complexity and delay characteristics with respect to each other.

\begin{table}[hbt!]
\vspace{-0.2cm}
\caption{Proposed Sorter Designs} 
\vspace{-0.2cm}
\centering 
\begin{tabular}{ >{\centering\arraybackslash} m{0.3in} | >{\centering\arraybackslash} m{0.5in} | >{\centering\arraybackslash} m{0.5in} | >{\centering\arraybackslash} m{0.5in} | >{\centering\arraybackslash} m{0.5in} } 
Design & Metric Sorting & Index Sorting & Complexity & Delay \\  [0.5ex] 
\hline
\hline 
 $1$ & MVF & Bitonic & Low & High \\   \hline
 $2$ & MVF & Radix-$L$ & Medium & Medium \\   \hline
 $3$ & Radix-$2L$ & Radix-$L$ & High & Low \\   \hline
\end{tabular}
\label{table:sorter_designs} 
\end{table} 

\section{Implementation Results}
\label{sec:results}
In this section, we verify the proposed sorting method by field programmable gate array (FPGA) implementations. 
We use Xilinx-XCZU9EG-FFVB1156-2-i in the implementations and the methods in \cite{chapman_mux} for obtaining multiplexers with number of inputs different from $2^{m}$.
\begin{table}[h!]
\vspace{-0.2cm}
\caption{LUT Consumptions of Conventional Crossbars (Synthesis Results)} 
\vspace{-0.2cm}
\centering 
\begin{tabular}{ >{\centering\arraybackslash} m{0.4in} | >{\centering\arraybackslash} m{0.6in} | >{\centering\arraybackslash} m{0.6in} | >{\centering\arraybackslash} m{0.6in}} 
  & $N=2048$ & $N=4096$ & $N=8192$   \\  [0.5ex] 
\hline
\hline 
 $L=4$ & $10140$ & $18948$ & $59113$   \\   \hline
 $L=8$ & $49600$ & $79133$ & $286330$  \\   \hline
 $L=16$ & $210572$ & $426616$ & $1437269$   \\   \hline
 $L=32$ & $852630$ & $1962224$ & $6610732$   \\   \hline
\end{tabular}
\label{table:crossbar_results} 
\vspace{-0.2cm}
\end{table} 

\begin{table*}[t!]
\vspace{-0.2cm}
\caption{Sorter Implementation Results} 
\centering 
\vspace{-0.2cm}
\begin{tabular}{ >{\centering\arraybackslash} m{0.4in} | >{\centering\arraybackslash} m{0.3in}| >{\centering\arraybackslash} m{0.25in}| >{\centering\arraybackslash} m{0.3in} | >{\centering\arraybackslash} m{0.25in} | >{\centering\arraybackslash} m{0.3in} | >{\centering\arraybackslash} m{0.25in} | >{\centering\arraybackslash} m{0.3in} | >{\centering\arraybackslash} m{0.25in} | >{\centering\arraybackslash} m{0.3in} | >{\centering\arraybackslash} m{0.25in} | >{\centering\arraybackslash} m{0.3in} | >{\centering\arraybackslash} m{0.25in} | >{\centering\arraybackslash} m{0.3in} | >{\centering\arraybackslash} m{0.25in}} 
  & \multicolumn{2}{c|}{SBS} & \multicolumn{2}{c|}{Radix-$2L$}& \multicolumn{2}{c|}{MVF} & \multicolumn{2}{c|}{OES} & \multicolumn{2}{c|}{Design 1} & \multicolumn{2}{c|}{Design 2} & \multicolumn{2}{c}{Design 3} \\  [0.5ex] 
\cline{2-15}
 & LUTs & Freq. (MHz) & LUTs & Freq. (MHz) & LUTs & Freq. (MHz) & LUTs & Freq. (MHz)& LUTs & Freq. (MHz)& LUTs & Freq. (MHz) & LUTs & Freq. (MHz)   \\  [0.5ex] 
\hline
\hline 
 $L=4$ & $124$ & $408.2$ & $351$ & $434.8$ & $322$ & $285.7$ & $295$ & $238.1$ & $390$ & $200.0$ & $443$ & $206.2$ & $461$ & $333.3$  \\   \hline
 $L=8$ & $589$ & $161.3$ & $1551$ & $285.7$ & $1073$ & $149.3$ & $526$ & $128.2$ & $1248$ & $87.7$ & $1671$ & $107.5$ & $2143$ & $188.7$  \\   \hline
 $L=16$ & $2982$ & $82.0$ & $6852$ & $172.4$ & $4141$ & $83.3$ & $1983$ & $74.6$ & $4579$ & $47.4$ & $5392$ & $62.5$ & $9246$ & $111.1$  \\   \hline
 $L=32$ & $13757$ & $32.1$ & $29541$ & $113.6$ & $11114$ & $54.6$ & $5949$ & $44.2$ & $17059$ & $29.0$ & $23375$ & $37.3$ & $43638$ & $66.2$  \\   \hline
\end{tabular}
\label{table:sorter_results} 
\vspace{-0.3cm}
\end{table*} 

First, we investigate the hardware complexities of the crossbars. 
Table~\ref{table:crossbar_results} presents the look-up table (LUT) numbers for conventional crossbars for different list sizes and bit widths. 
One can observe that significant numbers of LUTs are required for a single crossbar when the bit width is large.  

Table~\ref{table:sorter_results} gives the implementation results for state-of-the-art sorters and the sorter designs explained in the previous section. We consider the simplified bubble sorter (SBS) \cite{stimming_bubble}, radix-$2L$ sorter \cite{stimming_arch}, MVF \cite{lin_efficient} and simplified odd-even sorter (OES) \cite{oes} for comparison. From the results in Tables~\ref{table:crossbar_results}~and~\ref{table:sorter_results}, one can directly observe that the crossbar complexities are significantly higher than those of the sorters. 
Therefore, the increase in overall decoder complexity due to the employed sorter design is expected to be negligible compared to the gains obtained from crossbars with the proposed method. 

\begin{table}[h!]
\caption{Estimated LUT Gains from Crossbars} 
\centering 
\vspace{-0.2cm}
\begin{tabular}{ >{\centering\arraybackslash} m{0.4in} | >{\centering\arraybackslash} m{0.6in} | >{\centering\arraybackslash} m{0.6in} | >{\centering\arraybackslash} m{0.6in}} 
  & $N=2048$ & $N=4096$ & $N=8192$   \\  [0.5ex] 
\hline
\hline 
 $L=4$ & $2535$ & $4737$ & $14778$   \\   \hline
 $L=8$ & $18600$ & $29675$ & $107374$  \\   \hline
 $L=16$ & $92125$ & $186645$ & $628805$   \\   \hline
 $L=32$ & $399670$ & $919793$ & $3098781$   \\   \hline
\end{tabular}
\label{table:crossbar_gains} 
\vspace{-0.7cm}
\end{table} 

Secondly, one can observe that the proposed sorter designs offer a trade-off between hardware complexity and throughput, even though the complexity gain from crossbars is expected to be more significant, as mentioned above. 

Finally, comparing the state-of-the-art and the proposed sorters, one can observe that the proposed sorting method does not necessarily imply higher sorter complexity or delay. For example, the hardware consumption of a Radix-$2L$ sorter is higher than those of Designs $1$ and $2$ for $L > 4$ and $L > 8$, respectively. Design $3$ achieves higher throughput than those of SBS, MVF and OES for $L > 4$. Furthermore, there are large variations in hardware consumptions and operating frequencies also among the state-of-the-art sorters. We can conclude that the proposed sorting method can be implemented with lower hardware complexity or higher throughput than those of state-of-the art sorting methods depending on the design.   

Table~\ref{table:crossbar_gains} presents the estimated LUT gains from the crossbar implementations when the proposed sorting method is used. Comparing Tables~\ref{table:sorter_results}~and~\ref{table:crossbar_gains}, one can observe that the expected hardware consumption gains are much larger than possible hardware consumption increases due to the sorter designs in a decoder.   

\begin{table}[ht!]
	\vspace{-0.3cm}
	\caption{Decoder Implementation Results} 
	\centering
	\vspace{-0.2cm}
	\begin{tabular}{c|c|c|c|c}
		Decoder 		& \multicolumn{2}{c|}{Conv. w/ OES} & \multicolumn{2}{c}{Simp. w/ Design 3} \\ \hline  \hline
		$L$       	& 4 & 8 & 4 & 8 \\ \hline 
		$(N,K)$ 		& \multicolumn{4}{c}{($4096$, Any)} \\ \hline
		LUTs       	& $76229$ & $198082$ & $69058$  & $151901$    \\ \hline
		Registers		& $27445$ & $54371$ & $27392$ & $54371$   \\ \hline
		RAM [Mbits]  & $0.74$ & $1.46$ & $0.74$ & $1.46$  \\ \hline
		$f_{op}$[MHz] & $129.9$ & $66.7$ & $172.4$ & $96.2$  \\ \hline
		Latency [clock cycles]  & \multicolumn{4}{c}{$12928$} \\ \hline
		$TP$ [Mbps] & $41.1$ & $21.1$ & $54.6$ & $30.5$  \\ \hline
		Sorter $f_{op}$[MHz] & $238.1$ & $128.2$ & $333.3$ & $188.7$  \\ \hline
		Sorter LUTs & $295$ & $526$ & $461$ & $2143$  \\ \hline
	\end{tabular}
	\label{table:decoder_results}
\vspace{-0.2cm}
\end{table}

In order to validate the estimated gains in Table~\ref{table:crossbar_gains}, we implement SCL decoders with and without the proposed method. We use the semi-parallel architecture in \cite{asemiparallel} with $P=32$ processing elements and LLR-based metric calculation in \cite{stimming_llr}. Decoded bits are stored in registers of $N$ bits.
Calculated LLRs are stored in RAM blocks and pointer registers of $(\log_2{N}-1)\log_2{L}$ bits are used for pointer-based copying. 
Partial-sums are calculated by the partial-sum network in \cite{anefficientpartsumnet} and stored in registers of $P$ (parallel part) and $N/2$ (serial part) bits. LLRs are represented in the conventional sign-magnitude form, however, different LLR representation methods, such as \cite{yoon_llr}, can also be employed. All registers in the implementations are copied by crossbars with the corresponding input widths. The implemented decoders are flexible in terms of code rate. Conventional decoders without proposed sorting method employ simplified OES sorter, which is a low-complexity state-of-the-art sorter as seen from Table~\ref{table:sorter_results}. Implementation results are given in Table~\ref{table:decoder_results}. 

The results show that the proposed method achieves significant hardware consumption gain of SCL decoders. For the considered decoder architecture and block lengths, we obtain a hardware consumption gains of approximately $9.41\%$ for $L=4$ and $23.3\%$ for $L=8$ with the proposed method. The complexity gains can be verified from the crossbar implementation results in Table~\ref{table:crossbar_gains} for a $4096$-bit crossbar to copy decoded bit registers and a $2048$-bit crossbar to copy partial-sum registers. 

\begin{table}[h!]
	\vspace{-0.3cm}
	\caption{Comparison with State-of-the-Art Decoders} 
	\centering
	\vspace{-0.2cm}
	\begin{tabular}{ >{\centering\arraybackslash} m{0.75in} |  >{\centering\arraybackslash} m{0.5in} | >{\centering\arraybackslash} m{0.34in} | >{\centering\arraybackslash} m{0.34in} | >{\centering\arraybackslash} m{0.34in}}
		Decoder 		& Simp. w/ Design 3  & \cite{xiong_emulation} & \cite{liang_distributed} & \cite{xia_largelist} \\ \hline  \hline
		$L$       	& 8 & 4 & 4 & 32 \\ \hline 
		$(N,K)$ 	  & ($4096$, $2048$) & ($1024$, $512$) & ($1024$, $512$) & ($4096$, $2048$) \\ \hline
		LUTs/ALMs   & $153856$ & $142961$ & $101160$  & $67211$    \\ \hline
		Registers		& $54371$ & $19795$ & $13544$ & $31247$   \\ \hline
		RAM [Mbits]& $1.49$ & $4404$ & $0$ & $22440$  \\ \hline
		$f_{op}$[MHz] & $96.0$ & $42.66$ & - & $107$  \\ \hline
		Latency [cycles] & $7297$ & $371$ & $4064$ & $16019$  \\ \hline
		$TP$ [Mbps] & $54$ & $115$ & - & $27.35$  \\ \hline
	\end{tabular}
	\label{table:decoder_results}
\vspace{-0.2cm}
\end{table}

As specified in Table~\ref{table:crossbar_results}, the decoding latencies of the implemented decoders are equal to the latency of a semi-parallel SC decoder with additional $N$ clock cycles for path pruning operations. The sorting method we propose does not increase the decoding latency, thus the throughput values of the architectures in comparison have identical relations with the maximum achievable operation frequencies. The maximum delay path of the decoder passes through the sorter, as pointed out in \cite{stimming_llr}. We observe that the throughput of the decoders using the proposed sorting method are higher than those of the conventional decoders, owing to the smaller delay of Design $3$ sorter with respect to the delay of OES sorter. 
The results show that the decoders using proposed sorting method with proper sorter designs can also achieve significant gains in hardware consumption and higher throughput values with respect to those of decoders using conventional state-of-the-art sorters. 

Finally, we compare a decoder implementation employing the proposed sorting method with state-of-the-art SCL decoders. Table~\ref{table:decoder_results} gives the implementation results. We optimize our decoder for code rate $0.5$ for a fair comparison with the presented decoders. For this purpose, we apply certain simplifications of simplified successive cancellation list (SSCL) decoding \cite{hashemi_sscl}. Specifically, we perform rate-$0$ and repetition code simplifications for constituent codes of length up to $16$. For a polar code with $N=4096$ and $K=2048$, this reduces the decoding latency down to $7297$ with a similar hardware complexity, as also stated out in \cite{zhang_ssc}. We note that the throughput can further be improved by the optimizations in \cite{zhou_llr} without any change in the sorter architecture.

Compared with the decoders in \cite{xiong_emulation} and \cite{liang_distributed}, the decoder with the proposed sorting method has a significant advantage in terms of hardware complexity. More specifically, the proposed method can support twice the list size, which requires approximately four times as large hardware resources as observed from the results in Table~\ref{table:crossbar_results}, for a four times as large block length with similar hardware resources and throughput. The decoder in \cite{xia_largelist} achieves hardware complexity reduction by completely eliminating decoded bit and partial-sum crossbars at the expense of increased latency. Therefore, the decoder in \cite{xia_largelist} has a higher latency even though it benefits from multibit decoding (MBD) \cite{yuan_multibit} and switches to parallel decoding at low decoding stages to achieve latency reduction.
As seen from the implementation results, the simplified decoder with the proposed sorting method can achieve larger throughput than that of \cite{xia_largelist} with significantly lower memory consumption. On the other hand, the decoder in \cite{xia_largelist} can operate with larger list sizes owing to the lack of crossbars with large input widths. The results show that the proposed method offers a balanced decoder design with reasonable hardware complexity and throughput, especially for large block lengths. 

We note that the presented sorters and decoder are example implementations to verify the gains obtained by the proposed sorting method. The method can be applied in different decoder architectures and using different sorter designs. 
 
\section{Conclusion}
\label{sec:conclusion}
\vspace{-0.15cm}
In this work, a hardware complexity reduction method for SCL decoder implementations is proposed. The method comprises applying a sorting mechanism to candidate path metrics so that surviving paths are sorted according to their path indexes. With the proposed method, $(L/2+1)$-to-$1$ multiplexers can be employed instead of $L$-to-$1$ multiplexers for memory copying operations after path pruning. Implementation results show that significant reduction in hardware consumption is achievable with the proposed method without any penalty in decoding latency. Future work includes novel sorter designs and ASIC implementations for the proposed method.

\ifCLASSOPTIONcaptionsoff
  \newpage
\fi


\vspace{-0.3cm}
\begin{thebibliography}{10}
\providecommand{\url}[1]{#1}
\vspace{-0.1cm}
\bibitem{arikan}
E.~Ar{\i}kan, ``Channel polarization: a method for constructing
  capacity-achieving codes for symmetric binary-input memoryless channels,''
  \emph{IEEE Trans. Inform. Theory}, vol.~55, no.~7, pp. 3051--3073, July 2009.

\bibitem{tal_list}
I.~Tal and A.~Vardy, ``List decoding of polar codes,'' in \emph{Proc. IEEE Int.
  Symp. Inform. Theory (ISIT)}, July 2011, pp. 1--5.
	

\bibitem{DumerList} I. Dumer and K. Shabunov, ``Soft-decision decoding of Reed-Muller codes: recursive lists,'' 
IEEE Trans. Inform. Theory, vol. 52, no. 3, pp. 1260ֱ266, Mar. 2006.

						
\bibitem{stimming_llr}
A.~Balatsoukas-Stimming, M.~B.~Parizi and A.~Burg, ``{LLR}-based successive cancellation list decoding of polar codes,'' \emph{IEEE Trans. Signal  Process.}, vol.~63, no.~19, pp. 5165--5179, Oct.~2015.
	
\bibitem{stimming_arch}
A.~Balatsoukas-Stimming, A.~J.~Raymond, W.~J.~Gross and A.~Burg, ``Hardware architecture for list successive cancellation decoding of polar codes,'' \emph{IEEE Trans. Circuits and Syst. II, Exp. Briefs}, vol.~61, no.~8, pp.~609--613, Aug.~2014.	

\bibitem{yuan_multibit}
B.~Yuan and K.~K.~Parhi, ``Low-latency successive-cancellation list decoders for polar codes with multibit decision,'' \emph{IEEE Trans. Very Large Scale Integration (VLSI) Syst.}, vol.~23, no.~10, pp.~2268--2280, Oct.~2015.

\bibitem{fan_doublethr}
Y. Fan et al., ``Low-latency list decoding of polar codes with double thresholding,'' in \emph{2015 IEEE Intern. Conf. on Acoust., Speech and Signal Process. (ICASSP)}, Brisbane, QLD, 2015, pp.~1042--1046.

\bibitem{fan_lowlatency}
Y.~Fan et al., ``A low-latency list successive-cancellation decoding implementation for polar codes,'' \emph{IEEE Journ. on Sel. Areas in Commun.}, vol.~34, no.~2, pp.~303--317, Feb.~2016.

\bibitem{hashemi_parititoned}
S.~A.~Hashemi, A.~Balatsoukas-Stimming, P.~Giard, C.~Thibeault and W.~J.~Gross, ``Partitioned successive-cancellation list decoding of polar codes,'' in \emph{2016 IEEE Intern. Conf. on Acoust., Speech and Signal Process. (ICASSP)}, Shanghai, 2016, pp.~957--960.

\bibitem{hashemi_sphere}
S.~A.~Hashemi, C.~Condo and W.~J.~Gross, ``A fast polar code list decoder architecture based on sphere decoding,'' \emph{IEEE Trans. Circuits Syst. I, Reg. Papers}, vol.~63, no.~12, pp.~2368--2380, Dec.~2016.

\bibitem{hashemi_fast}
S.~A.~Hashemi, C.~Condo and W.~J.~Gross, ``Fast and Flexible Successive-Cancellation List Decoders for Polar Codes,'' \emph{IEEE Trans. Signal  Process.}, vol.~65, no.~21, pp.~5756--5769, Nov.~2017.

\bibitem{sural_thesis}
A.~Sural, ``An FPGA implementation of successive cancellation list decoding for polar codes,'' Bilkent University masters
thesis, Feb. 2016.

\bibitem{song_efficient}
W.~Song, C.~Zhang, S.~Zhang and X.~You, ``Efficient adaptive successive cancellation list decoders for polar codes,'' in \emph{2016 IEEE Int. Conf. on Digital Signal Process. (DSP)}, Beijing, 2016, pp.~218--222.

\bibitem{yuan_multibit}
B.~Yuan and K.~K.~Parhi, ``{LLR}-based successive-cancellation list decoder for polar codes with multibit decision," \emph{IEEE Trans. Circuits and Syst. II, Exp. Briefs}, vol.~64, no.~1, pp.~21--25, Jan.~2017.

\bibitem{zhang_hardware}
C.~Zhang, X.~You and J.~Sha, ``Hardware architecture for list successive cancellation polar decoder,'' in \emph{2014 IEEE Int. Symp. Circuits Syst. (ISCAS)}, Melbourne VIC, 2014, pp.~209--212.

	
\bibitem{lin_hightp}
J.~Lin, C.~Xiong and Z.~Yan, ``A high throughput list decoder architecture for polar codes,'' \emph{IEEE Trans. Very Large Scale Integration (VLSI) Syst.}, vol.~24, no.~6, pp.~2378--2391, June~2016.

\bibitem{xiong_multimode}
C.~Xiong, J.~Lin and Z.~Yan, ``A multimode area-efficient {SCL} polar decoder,'' \emph{IEEE Trans. on Very Large Scale Integration (VLSI) Syst.}, vol.~24, no.~12, pp.~3499--3512, Dec.~2016.

\bibitem{xia_onpath}
C.~Xia, Y.~Fan, J.~Chen and C.~Tsui, ``On path memory in list successive cancellation decoder of polar codes,'' in \emph{2018 IEEE Int. Symp. Circuits Syst. (ISCAS)}, Florence, Italy, 2018, pp.~1--5.

\bibitem{xiong_emulation}
C.~Xiong, Y.~Zhong, C.~Zhang and Z.~Yan, ``An {FPGA} emulation platform for polar codes,'' \emph{2016 IEEE Int. Workshop on Signal Process. Syst. (SiPS)}, Dallas, TX, 2016, pp.~148--153.

\bibitem{lin_efficient}
J.~Lin and Z.~Yan, ``An efficient list decoder architecture for polar codes,'' \emph{2016 IEEE Trans. on Very Large Scale Integration (VLSI) Syst.}, vol.~23, no.~11, pp.~2508--2518, Nov.~2015.

\bibitem{xiong_symbol}
C.~Xiong, J.~Lin and Z.~Yan, ``Symbol-decision successive cancellation list decoder for polar codes,'' \emph{IEEE Trans. Signal  Process.}, vol.~64, no.~3, pp.~675--687, Feb.~2016.

\bibitem{yuan_llr}
B.~Yuan and K.~K.~Parhi, ``Successive cancellation list polar decoder using log-likelihood ratios,'' in \emph{Proc. IEEE Asilomar Conf. Signals, Syst., Comput.}, 2014, pp.~548–-552.

\bibitem{radix_masera}
L.~G.~Amaru, M.~Martina and G.~Masera, ``High speed architectures for finding the first two maximum/minimum values,'' \emph{IEEE Trans. on Very Large Scale Integration (VLSI) Systems}, vol.~20, no.~12, pp.~2342--2346, Dec.~2012.

\bibitem{batcher_sorting}
K.~E.~Batcher, ``Sorting networks and their applications,'' in \emph{Proc. Spring Joint Comput. Conf.}, Atlantic City, NJ, USA, May~1968, pp.~307--314.

\bibitem{chapman_mux}
K.~Chapman, ``Multiplexer design techniques for datapath performance with minimized routing resources,'' Oct.~31, 2014, XILINX.

\bibitem{stimming_bubble}
A.~Balatsoukas-Stimming, M.~Bastani~Parizi and A.~Burg, ``On metric sorting for successive cancellation list decoding of polar codes,'' \emph{2015 IEEE Int. Symp. Circuits Syst. (ISCAS)}, Lisbon, 2015, pp.~1993--1996.

\bibitem{oes}
B.~Yong~Kong, H.~Yoo and I.~Park, ``Efficient sorting architecture for successive-cancellation-list decoding of polar codes,'' \emph{IEEE Trans. Circuits and Syst. II, Exp. Briefs}, vol. 63, no. 7, pp. 673-677, July 2016.

\bibitem{asemiparallel}
C.~Leroux, A.~Raymond, G.~Sarkis, and W.~Gross, ``A semi-parallel successive-cancellation decoder for polar codes,'' \emph{IEEE Trans. Signal Process.}, vol.~61, no.~2, pp.~289--299, Jan.~2013.
	
\bibitem{anefficientpartsumnet}
Y.~Fan and C.-Y. Tsui, ``An efficient partial-sum network architecture for semi-parallel polar codes decoder implementation,'' \emph{IEEE Trans. Signal Process.}, vol.~62, no.~12, pp. 3165--3179, June 2014.
			
\bibitem{yoon_llr}
H.~Yoon~and~T.~Kim, ``Efficient successive-cancellation polar decoder based on redundant {LLR} representation,'' \emph{ IEEE Trans. Circ. Sys. II: Exp. Briefs}, vol.~65, no.~12, pp.~1944--1948, Dec.~2018.

\bibitem{hashemi_sscl}
S.~A.~Hashemi, C.~Condo and W.~J.~Gross, ``Simplified successive-cancellation list decoding of polar codes,'' \emph{2016 IEEE Int. Symp. Inf. Theory (ISIT)}, Barcelona, 2016, pp. 815--819.

\bibitem{zhang_ssc}
C.~Zhang~and~K.~K.~Parhi, ``Latency analysis and architecture design of simplified {SC} polar decoders,'' \emph{ IEEE Trans. Circ. Sys. II: Exp. Briefs}, vol.~61, no.~2, pp.~115--119, Feb.~2014.

\bibitem{zhou_llr}
Y.~Zhou, Z.~Chen, J.~Lin and Z.~Wang, ``A high-speed successive-cancellation decoder for polar codes using approximate computing,'' \emph{ IEEE Trans. Circ. Sys. II: Exp. Briefs}, vol.~66, no.~2, pp.~227--231, Feb.~2019.

\bibitem{liang_distributed}
X.~Liang, C. Yang, J.and Zhang, W. Song, and X. You, ``Hardware efficient and low-latency {CA-SCL} decoder based on distributed sorting,'' \emph{2016 IEEE Glob. Commun. Conf. (GLOBECOM)}, Washington, DC, 2016, pp. 1-6.

\bibitem{xia_largelist}
C.~Xia et al., ``An implementation of list successive cancellation decoder with large list size for polar codes,'' \emph{2017 27th Int. Conf. on Field Prog. Logic and App. (FPL)}, Ghent, 2017, pp. 1-4.




\end{thebibliography}
\end{document}